# Disentanglement of mixed interference fringes in optical interferometers: theory and applications

KAIYUAN YANG,[1,2,3,4] WEILONG WEI,[1,3,4] XIAFEI MA,[1,3,4] BOTAO CHEN,[1,3,4] JUNQIU CHU,[1,3,4] XINLING LIU, [1,3,4] YUHUA CHENG,[2] HU YANG, [1,3,4] HAOTONG MA,[1,3,4,5] BO QI,[1,3,4,6] AND ZONGLIANG XIE,[1,3,4, *]

[1]*Institute of Optics and Electronics, Chinese Academy of Sciences, Chengdu 610209, China*
[2]*University of Electronic Science and Technology of China, Chengdu 610054, China*
[3]*University of Chinese Academy of Sciences, Beijing 100039, China*
[4]*Key Laboratory of Optical Engineering, Chinese Academy of Sciences, Chengdu 610209, China*
[5]*mahaotong@163.com*
[6]*qibo@ioe.ac.cn*
[*]*xiezl@ioe.ac.cn*

**Abstract:** Optical interferometric imaging enables astronomical observation at extremely high angular resolution. The necessary optical information for imaging, such as the optical path differences and visibilities, is easy to extract from fringes generated by the combination of two beams. With more than two apertures, the image-plane interference pattern becomes an increasingly indistinguishable mixture of fringe spacings and directions. For decades, the state-of-the-art approaches for obtaining two-aperture fringes from an interferometer array composed of many apertures are limited to pairwise combinations using bulk optics. Here, we derive and demonstrate a fringe disentanglement theory that can digitally transform the interference pattern of $N$ apertures to $N(N-1)/2$ pairwise fringes without any optics, thus providing straightforward methods of information acquisition for interferometers. We demonstrate applications of our technique by both simulation and experiment, showing that this theory can be used for simultaneously sensing pistons and determining the individual visibilities of all combining apertures. Furthermore, we use the proposed theory to phase a 1.5-meter segmented flat telescope, demonstrating its validity for engineering implementation. This theory may not only benefit optical imaging but also interferometry-based measurements, by providing an exceptional capability to simplify the interferometric output generated by a system of many apertures.

## 1. Introduction

Optical interferometric imaging can achieve high spatial resolution down to milliarcseconds for astrophysical studies by combining light from multiple subapertures [1-7]. There are two main types of interferometric imaging modes in terms of beam combination schemes. The first is called the Fizeau configuration, in which the exit pupils are downscaled replicas of the input pupils [8], which can directly produce useful images via image-plane beam combinations, like large telescopes with masked apertures. The Fizeau interferometer is characterized by short baselines. In the extreme case where the distance between every two adjacent subapertures is zero, Fizeau interferometers, such as Keck [9-11] and JWST [12,13], are known as segmented telescopes [14,15]. The second is called the Michelson configuration, which is applied to interferometers without homothetic mapping of the exit or input pupils [16,17]. Michelson interferometers, such as VLTI [18-20] and CHARA [21-23], transport light from two or more separated telescopes with long baselines to the pupil-plane [24-26] or image-plane [27-29] beam combiners to produce interference fringes, from which the visibilities are measured and subsequently used to form images through aperture synthesis.

For both interferometric imaging schemes, it is important to retrieve underlying information from the interference patterns. The precise extraction of interferometric observables, such as optical path differences and visibilities, can be performed by processing a single fringe resulting from a two-beam combination. However, regardless of the interferometric scheme used, combining beams of more than two apertures to the same focus results in spatial accumulation of each set of fringes of each pupil pair on the image plane. The resulting speckle-like interference pattern loses fringe contrast, spacing and direction, thus leading to problems in information acquisition.

For decades, the state-of-the-art approaches for acquiring pairwise fringes from an interferometric array consisting of more than two units rely on the use of complex optical components. Beam splitters are common components used in optical interferometers for realizing pairwise interference [30-39]. An increase in the number of apertures would make the beam splitting device more complex. Keck segmented telescope utilizes a prism array to obtain well-separated pairwise interference patterns generated by combination of adjacent sub-aperture edge beams [40,41]. The currently largest space telescope, JWST, employs a specially designed dispersing element that spans adjacent sub-mirrors for spectral dispersion, thus producing dual-aperture interferometric dispersed fringes separated in the focal plane [42,43]. The similar dispersing hardware is also adopted in segmented Giant Magellan Telescope (GMT) to obtain isolated two-aperture dispersed fringes [44,45]. Recently, a Mach–Zehnder interferometry-based module composed of a lenslet array and prisms has been proposed for simultaneous acquisition of four pairwise combination spots on the detector [46].

Instead of complex physical solutions, this paper proposes a mathematic insight into disentanglement of mixed interference patterns, i.e., digitally decoupling pairwise fringes from an interference pattern resulting from combination of more than two sub-apertures. The theorical model of fringe disentanglement based on linear projection integration is established through mathematical derivation. We rigorously demonstrate that the projection integration of the mixed interference pattern along one selected baseline direction can be approximated as a combination of the dual-aperture fringes corresponding to the same baseline and incoherent accumulation of diffraction intensity of the remaining apertures. Eliminating the irrelevant analytical intensity distribution from the projection integration signal allows for the accurate extraction of pairwise interference fringes from the mixed speckle patterns.

This theory provides great benefits for optical interferometry. Here we present two applications of the fringe disentanglement theory in the field of interferometric imaging. One is piston sensing for multiple apertures. Thanks to fringe decoupling, we can perform multiple channel fringe tracking, thus achieving large range and high accuracy piston sensing. The disentanglement theory-based phasing technique is applicable to multiple apertures without any additional optical elements, and its capture range is limited by only the modulation range, not by the coherence length of light. Furthermore, we utilize the method to phase the 1.5-meter segmented flat telescope, demonstrating its validity for engineering implementation. The other is parallel measurement of the individual visibility. For all-in-one combination, currently, the only way to extract the visibilities resulting from each combination pair is to perform Fourier transform and discern the individual visibilities in the image intensity spectrum. The scenario that the signals from different baselines overlap in the frequency domain can result in a systematic error named crosstalk [47]. The fringe disentanglement theory offers an alternative solution for determination of visibilities of individual baselines in spatial domain, which can effectively reduce crosstalk with higher computational efficiency. We believe that the proposed theory is useful not only for interferometric imaging but also for making a positive impact on interferometry-based measurement [48-51].

## 2. Theory

### 2.1 Optical model of imaging interferometry

The angular resolution of imaging interferometer is determined by the longest baselines recorded through the combination of individual telescopes. For good image quality, existing interferometers have reached baseline lengths of hundreds of meters. Moreover, many observations at baselines for different length and orientation need to be made. For Fizeau configuration, the entrance-pupil and the exit-pupil are downscaled. For Michelson configuration, because most individual telescope with aperture diameter much smaller than their baseline, $\delta$-function is used to define the pinhole size of subaperture and the entrance-pupil can be expressed as:

$$A_N(\xi,\zeta) = \delta(\xi-\xi_{p_1},\zeta-\zeta_{p_1}) + \sum_{i=2}^{N} \delta(\xi-\xi_{p_i},\zeta-\zeta_{p_i})\exp(j\varphi_i) \tag{1}$$

where $N$ represents the number of subapertures, $(\xi,\zeta)$ are the coordinates of the entrance-pupil plane and $(\xi_{p_i},\zeta_{p_i})$ are the positions of the $n$th subaperture, and $\varphi_i$ is the relative wavefront error between the $n$th nonreference subaperture and the reference. In contrast to the Fizeau configuration, the light received from several separate telescopes is input into the beam combination (BC) instrument for interference, which is chosen independently of the interferometer baseline. Consequently, the exit-pupil function should be expressed as follows:

$$P_N(\mu,\nu) = p(\mu-\mu_{p_1},\nu-\nu_{p_1}) + \sum_{i=2}^{N} p(\mu-\mu_{p_i},\nu-\nu_{p_i})\exp(j\varphi_i))$$

$$p(\mu,\nu) = \begin{cases} 1 & \text{in the subpupil} \\ 0 & \text{out of subpupil} \end{cases} \tag{2}$$

where $(\mu,\nu)$ represents the coordinates of the exit-pupil plane, $(\mu_{p_i},\nu_{p_i})$ is the center position of the $n$th subaperture in the exit-pupil plane, and $p(\mu,\nu)$ is the binary pupil function of an elementary. Based on the Fourier imaging principle, especially in cases of incoherent and monochromatic signals, the point spread function (PSF) of the interferometric imaging system is determined by the exit-pupil function, which can be expressed as:

$$PSF_N(x,y,\lambda) = \left|FT\{P_N(\mu,\nu)\}\right|^2 = 2 * \frac{\frac{D^2}{2} f^2 \lambda^2 J_1^2(\pi D\sqrt{x^2+y^2})}{x^2+y^2} \left\{\sum_{i=1}^{N} \exp\left[-\frac{2\pi j}{\lambda f}\left(x\mu_{p_i}+y\nu_{p_i}\right)+j\varphi_i\right]\right\}^2$$
$$= \mathbf{F}_{diffraction} * \mathbf{F}_{interference} \tag{3}$$

where $(x,y)$ are the coordinates of the image plane, $J_1$ represents the first-order Bessel function and $D$ is the diameter of the subaperture. For convenient analysis, the PSF can be defined as the product of the diffraction factor and interference factor. The high-frequency component of a PSF image is solely determined by the interference factor, while the diffraction factor serves as an envelope for the PSF. Therefore, the PSF of imaging interferometry can be further simplified as follows:

$$PSF_N(x,y,\lambda) = PSF_{sub}(x,y,\lambda) * (n + 2\sum_{j=1}^{n-1}\sum_{k=j+1}^{n} \cos\left[(\mu_{p_j}-\mu_{p_k})\frac{2\pi x}{\lambda f} + (\nu_{p_j}-\nu_{p_k})\frac{2\pi y}{\lambda f} + \varphi_{jk}\right] \tag{4}$$

where $PSF_{sub}$ represents the diffraction envelope generated by a single subaperture and $\varphi_{jk}$ is the wavefront error between the $j$th subaperture and $k$th subaperture.

*2.2 Theoretical Model of Fringe Disentanglement*

As shown in Fig.1, according to the optical model of imaging interferometry, the PSF speckle can be considered as the entanglement of fringe patterns with the contributions of all baselines, specifically an increasingly indistinguishable mix of fringe spacings and orientations. Consequently, when observing with more than two apertures, it is difficult to distinguish the features of each individual baseline from the PSF. Here we establish a theorical model of fringe disentanglement based on linear projection integration that can simultaneously decompose the fringe of each individual baseline in a given speckle pattern, the schematic of which is presented in Fig.1.

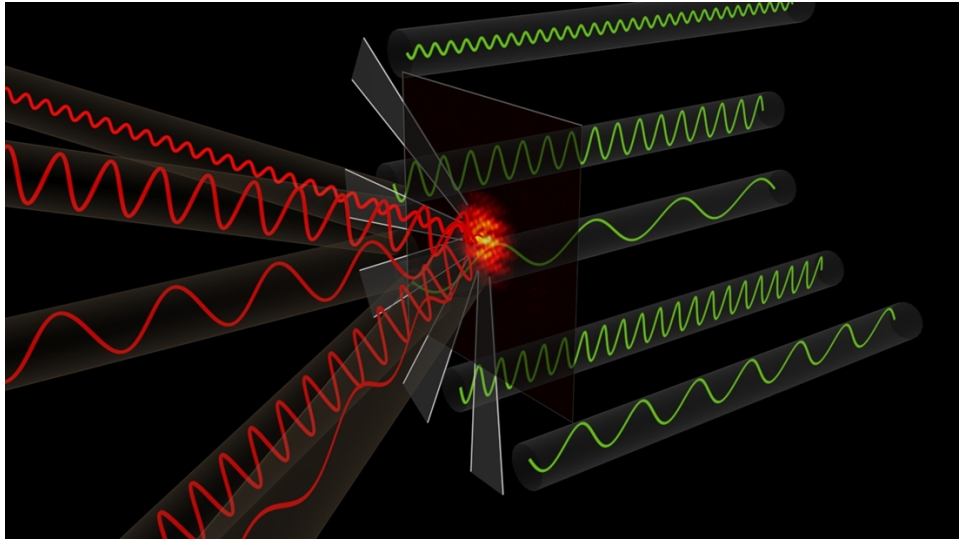

Fig. 1 Schematic diagram of fringe disentanglement theory.

By operating the linear projection integral (LPI) on the PSF image, the speckle pattern can be decomposed into a set of integral signals. All integral signals corresponding to each integration path in the projection space (0-π) compose the 2D integral image (LPI_IMA), which can be mathematically expressed as follows:

$$LPI\_IMA(\theta,b) = \int_{L(\theta,b)} PSF_{sub}(x,y,\lambda)\left\{n+2\sum_{j=1}^{n-1}\sum_{k=j+1}^{n}\cos\left[\left(\mu_{p_j}-\mu_{p_k}\right)\frac{2\pi x}{\lambda f}+\left(\nu_{p_j}-\nu_{p_k}\right)\frac{2\pi y}{\lambda f}\right]\right\}ds \quad (5)$$

s. t. $L(\theta,b): y = \tan(\theta)x + b$

where $L(\theta,b)$ is the path of integration, and $\tan\theta$ and $b$ represent the slope and intercept of the path, respectively. The projection signal can be classified as the first form of curve integral, where the integrand function is both continuous and differentiable. Consequently, the integral image can be simplified into a definite integral form:

$$LPI\_IMA(\theta,b) = \int_{x_{lb}}^{x_{ub}} PSF_{sub}\sqrt{1+(\tan\theta)^2}\left\{n+2\sum_{j=1}^{n-1}\sum_{k=j+1}^{n}\cos\left[\begin{array}{l}\left(\mu_{p_j}-\mu_{p_k}\right)\frac{2\pi x}{\lambda f}+\\+\left(\nu_{p_j}-\nu_{p_k}\right)\frac{2\pi(x\tan\theta+b)}{\lambda f}\end{array}\right]\right\}dx$$

s. t. $(x_{ub}, x_{lb}) \in \begin{cases} C: x^2+y^2=\rho^2 \\ L: y = x\tan\theta + b \end{cases}$

(6)

where $\rho$ is the radius of the Airy disk, and the variables $x_{ub}$ and $x_{lb}$ represent the upper and lower bounds of integral, which correspond to the intersection between the integration path and the edge of the Airy disk. Then, the interference pattern between subaperture 1 and subaperture 2 is taken as an illustrative example to demonstrate its extraction from LPI_IMA. First, the integration path is set orthogonal to the baseline, which is formed by subaperture 1 and subaperture 2, and the integrand function is written as the sum of the disentangled fringe and remaining fringes. The single linear projection integral signal can be expressed as:

$$LPI(b) = \sqrt{1+(\tan\theta)^2} \times \int_{x_{lb}}^{x_{ub}} PSF_{sub} \left\{ \begin{array}{l} 2+2\cos\left[(x_1-x_2)\dfrac{2\pi x}{\lambda f}+(y_1-y_2)\dfrac{2\pi(x\tan\theta+b)}{\lambda f}\right]+(n-2)+ \\ +\sum_{j=1}^{n-1}\sum_{k=j+1,k\neq 2}^{n} 2\cos\left[(x_j-x_k)\dfrac{2\pi x}{\lambda f}+(y_j-y_k)\dfrac{2\pi(x\tan\theta+b)}{\lambda f}\right] \end{array} \right\} dx \quad (7)$$

Because the product of the tangent value of the integration path and the baseline direction is zero, the integral equation is expressed as follows:

$$LPI(b) = \sqrt{1+(\tan\theta)^2} \int_{x_{lb}}^{x_{ub}} PSF_{sub} \left\{ \begin{array}{l} 2+2\cos\left[(y_1-y_2)\dfrac{2\pi b}{\lambda f}\right]+(n-2)+ \\ +\sum_{j=1}^{n-1}\sum_{\substack{k=j+1 \\ k\neq 2}}^{n}\cos\left[\begin{array}{l}(x_j-x_k)\dfrac{2\pi x(1+\tan\theta\tan\theta_{jk})}{\lambda f}+ \\ +(y_j-y_k)\dfrac{2\pi b}{\lambda f}\end{array}\right] \end{array} \right\} dx \quad (8)$$

where $\theta_{jk}$ represents the baseline direction between the $j$th subaperture and $k$th subaperture. Subsequently, the integral signal is simplified as the sum of three components:

$$LPI(b) = 2\sqrt{1+(\tan\theta)^2}\left\{1+\cos\left[\dfrac{2\pi b(y_1-y_2)}{\lambda f}\right]\right\}\int_{x_{lb}}^{x_{ub}} PSF_{sub}dx + (n-2)\sqrt{1+(\tan\theta)^2}\int_{x_{lb}}^{x_{ub}} PSF_{sub}dx +$$

$$+\sum_{j=1}^{n-1}\sum_{\substack{k=j+1 \\ k\neq 2}}^{n}\int_{x_{lb}}^{x_{ub}} PSF_{sub}\left\{\cos\left[(x_j-x_k)\dfrac{2\pi x(1+\tan\theta\tan\theta_{jk})}{\lambda f}+(y_j-y_k)\dfrac{2\pi b}{\lambda f}\right]\right\}dx \quad (9)$$

$$=LPI_1+LPI_2+LPI_3$$

Here, we analyze the physical significance of these three components in the integrated signal and compare their values. The first term $LPI_1$ contains the fringe that requires disentanglement. The second term $LPI_2$ is regarded as the incoherent superposition of the remaining n-2 subapertures, while the third term $LPI_3$ displays the accumulated interference factors along the integration path between them. The value of the incoherent superposition term clearly increases with increasing number of subapertures, which is numerically analyzed and thus can be eliminated. When considering the narrow region around the central position of the Airy disk, $b$ can be approximated as 0, and the linear integral of $PSF_{sub}$ can thus be considered constant $\varepsilon$. Subsequently, the Airy spot radius formulation is employed to simplify the coefficient component of the third term in Eq. 9 while simultaneously plugging the upper and lower limits of the integral into the equation. Consequently, the ratio between the first and third parts of the integral signal can be expressed as follows:

$$\frac{LPI_3}{LPI_1} = \frac{\sum_{j=1}^{n-1}\sum_{\substack{k=j+1\\k\neq 2}}^{n}\int_{x_{lb}}^{x_{ub}} PSF_{sub}\left\{\cos\left[(x_j-x_k)\frac{2\pi x(1+\tan\theta\tan\theta_{jk})}{\lambda f}+(y_j-y_k)\frac{2\pi b}{\lambda f}\right]\right\}dx}{2\sqrt{1+(\tan\theta)^2}\left\{1+\cos\left[\frac{2\pi b(y_1-y_2)}{\lambda f}\right]\right\}\int_{x_{lb}}^{x_{ub}} PSF_{sub}dx}$$

$$= \frac{\sum_{j=1}^{n-1}\sum_{\substack{k=j+1\\k\neq 2}}^{n}\frac{C\rho D\varepsilon}{1.22\pi L_{jk}\cos\Delta\theta}\left\{2\sin\left[\frac{2.44\pi L_{jk}}{D}\cos\Delta\theta\right]\right\}}{8C\rho\varepsilon} \quad (10)$$

$$= \sum_{j=1}^{n-1}\sum_{\substack{k=j+1\\k\neq 2}}^{n}\frac{1}{2}\frac{\sin\left[\frac{2.44\pi L_{jk}}{D}\cos\Delta\theta\right]}{\frac{2.44\pi L_{jk}\cos\Delta\theta}{D}} = \sum_{j=1}^{n-1}\sum_{\substack{k=j+1\\k\neq 2}}^{n}\frac{1}{2}\frac{\sin\psi}{\psi} = \sum_{j=1}^{n-1}\sum_{\substack{k=j+1\\k\neq 2}}^{n} f(\psi)$$

s. t. $\psi = \frac{2.44\pi L_{jk}}{D}\cos\Delta\theta$

where $D$ represents the diameter of the subaperture, $L_{jk}$ is the baseline length between the $j$th subaperture and $k$th subaperture, and $\Delta\theta$ denotes the angle that forms between the direction of $L_{jk}$ and the integration path. The range of the ratio can be determined by analyzing the characteristics of the function $f(\psi)$, as can be considered the sum of $f(\psi)$. Function $f(\psi)$ clearly exhibits a cosine-like behavior with continuous decay; thus, the oscillation amplitude of the function can be determined by its extreme point, which is calculated by the gradient of the function. This extreme point can be expressed in the form of a transcendental equation:

$$grad(f(\psi)) = \frac{\psi\cos\psi - \sin\psi}{2\psi^2} = 0 \quad (11)$$

The solution to the transcendental equation can be asymptotically expressed as:

$$\psi = k\pi + \arctan\left(k\pi + \frac{\pi}{2}\right) + (\varepsilon + \varepsilon^2)\left[\arctan\left(k\pi + \frac{\pi}{2}\right) - \frac{\pi}{2}\right] \quad \varepsilon = \left[1+\left(k\pi + \frac{\pi}{2}\right)^2\right]^{-1} \quad (12)$$

s. t. $k = 1, 2....n$

Certain extreme points and their corresponding function extremums are analyzed in Eq. (13). The value of the third term in the integrated signal decreases to less than one hundredth of the first term when $\psi$ is larger than 45.5. This constraint can be ensured by increasing the ratio between the baseline length and the diameter of the subaperture in the exit pupil plane. Thus, $LPI_3$ can be completely neglected as the noise component in $LPI$.

$$\begin{aligned}
\psi &= 7.7253, & f(\psi) &= 0.0642\\
\psi &= 14.0662, & f(\psi) &= 0.0354\\
\psi &= 20.3713, & f(\psi) &= 0.0245\\
\psi &= 32.9564, & f(\psi) &= 0.0151\\
\psi &= 45.5311, & f(\psi) &= 0.0109\\
&\vdots & &\vdots
\end{aligned} \quad (13)$$

Finally, the linear projection integral can be divided into three parts, namely, a decoupling term, an irrelevant term, and a noise term, which are expressed as follows:

$$LPI(b) = Dc(b,\theta,y_1,y_2) + Irr(n,b,\theta) + N(\rho,b,\theta_{jk},\theta,L_{jk})$$

$$\text{s. t. } Dc(b,\theta,y_1,y_2) = 2\varepsilon\sqrt{(1+(\tan\theta)^2)}\left\{1+\cos\left[\frac{2\pi b(y_1-y_2)}{\lambda f}\right]\right\}$$

$$Irr(n,b,\theta) = (n-2)\varepsilon\sqrt{(1+(\tan\theta)^2)} \tag{14}$$

$$N(\rho,b,\Delta\theta,\theta,L_{jk}) = \sum_{j=1}^{n-1}\sum_{\substack{k=j+1\\k\neq 2}}^{n}\int_{x_{lb}}^{x_{ub}} PSF_{sub}\left\{\cos\left[\begin{array}{l}(x_j - x_k)\dfrac{2\pi x(1+\tan\theta\tan\theta_{jk})}{\lambda f} + \\ +(y_j - y_k)\dfrac{2\pi b}{\lambda f}\end{array}\right]\right\}dx$$

Thus, the disentangled interference fringe can be expressed as follows:

$$Dc(b,\theta,y_1,y_2) \approx LPI(b) - Irr(n,b,\theta) \tag{15}$$

## 2.3 Numerical verification of the theory

To test the accuracy of the mathematical derivation, the numerical verification is performed. We take a simple three-aperture demonstration as an example. The analytic LPI intensity distribution of different baseline directions in a three-aperture system is obtained according to the formulas of Eq. (14) and then is compared with the numerical simulation results to verify the theorical accuracy.

The exit-pupil distribution of the three-aperture imaging system is shown in Fig. 2(a). The diameter of each exit-pupil is 15 mm, the equivalent focal length is 2 m, the pixel size of the detector is 1.67 μm, and the wavelength of monochromatic illumination is set to 500 nm. The PSF of the simulated imaging system is shown in Fig. 2(b).

According to the exit–pupil distribution, three projection integral paths corresponding to the baseline orientations can be obtained. Then the decoupling term and the irrelevant term of each LPI intensity distribution can be calculated from Eq. (14). Each intensity function was sampled at 1.67 μm intervals. The irrelevant terms $Irr(n,b,\theta)$ and decoupling terms $Dc(b,\theta,y_1,y_2)$ are computationally obtained as shown in Figs. 2(c1) - 2(c3) and Figs. 2(c4) - 2(c6), respectively. Consequently, by the Eq. (15), each analytical LPI can be approximately calculated by the sum of decoupling term and irrelevant term, as illustrated in Figs. 2(c7) - 2(c9). The corresponding three integral signals resulting from numerical simulations are presented in Figs. 2(d1) - (d3), and the residual errors between the analytic and simulated LPI results are shown in Figs. 2(d4) - 2(d6). From the comparison, it can be obviously found that the curves derived by Eq. (14) are nearly the same as the curves obtained by numerical simulation, verifying the theory of fringes disentanglement. The slight residuals should come from the analytic noise term of Eq. (14).

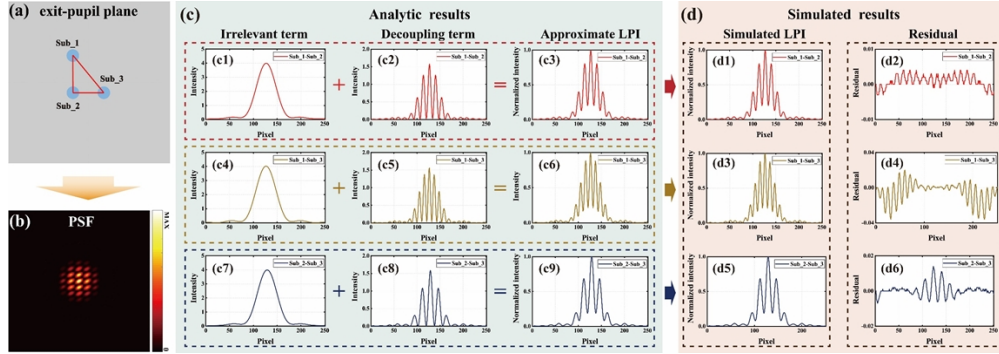

Fig. 2 Numerical verification of the fringe disentanglement theory. (a) Exit–pupil distribution. (b) The PSF of the simulated imaging system. (c) The analytic results using the theory of fringe

disentanglement, where (c1), (c4), and (c7) are the irrelevant terms of the LPI intensity and (c2), (c5), and (c7) are the decoupling terms of the LPI intensity. According to the Eq. (15), each analytically LPI can be approximately calculated by the sum of decoupling term and irrelevant term, as illustrated in Figs. 2(c3), 2(c6), and 2(c9), respectively. (d1) - (d3) The corresponding three integral signals resulting from numerical simulations, and (d4) - (d6) the residual errors between the analytic and simulated LPI results. It is obvious that the curves derived by Eq. (15) are nearly the same as the curves obtained by numerical simulation.

## 3. Application I - Piston Sensing

### 3.1 Piston sensing method

According to the theoretical analysis in Section 2.2, the fringe disentanglement theory based on the linear projection integral (LPI) can be applied to piston sensing, which requires decoupling only the interference fringes between the reference and nonreference subapertures from the PSF image. To eliminate 2π ambiguity, the Wide spectrum illumination is used and the decoupled fringe package naturally evolves into a linear superposition of fringe under multiple spectral channels, which are expressed as follows:

$$\sum_{i=1}^{n} \text{Dc}(\rho, b, \theta, y_1, y_2, \lambda_i) S(\lambda) \approx \sum_{i=1}^{n} [LPI(b, \lambda_i) - Irr(n, \rho, b, \theta, \lambda_i)] S(\lambda) \quad (16)$$
$$\text{s. t. } \Delta\lambda = \lambda_n - \lambda_1$$

where $\Delta\lambda$ represents the bandwidth, which is divided into n intervals equally and $S(\lambda)$ is the intensity weight of the fringe in different spectral channel, assuming $S(\lambda)=1$.

However, the primary mirrors of synthetic aperture imaging systems are mostly centered symmetric structures, and multiple pairs of parallel baselines lead to the same interference direction and thus are difficult to decouple on the image plane. Specifically, the fringe information of each parallel submirror pair with the same baseline direction is distributed in the same path of the LPI image. For example, in a four-aperture imaging system, the 1-2 subaperture pair and the 3-4 subaperture pair have the same interference direction; moreover, the 1-3 subaperture pair and the 2-4 subaperture pair also have the same interference direction, as illustrated in Fig. 3(b1). To eliminate the limitation of pupil arrangement, a mask without a parallel baseline was set on the exit-pupil plane to sample the wave reflected by the wavefront compensator, which is reflected on the entrance-pupil plane such that the mask is equivalent to a physical mask superimposed on the pupil, as shown in Fig. 2(a). This beam path is used to correct the pistons between subapertures. After the optical system is phased, the actual imaging beam path is still imaged by the segmented primary mirror. The two beam paths can be easily separated with a beam splitter. With the same four subaperture imaging systems, a special mask was designed to verify the decoupling effect on the interference fringe. Under these conditions, the 1-2, 1-3, and 1-4 subaperture pairs have different interference paths in the LPI image, as shown in Fig. 3(b2).

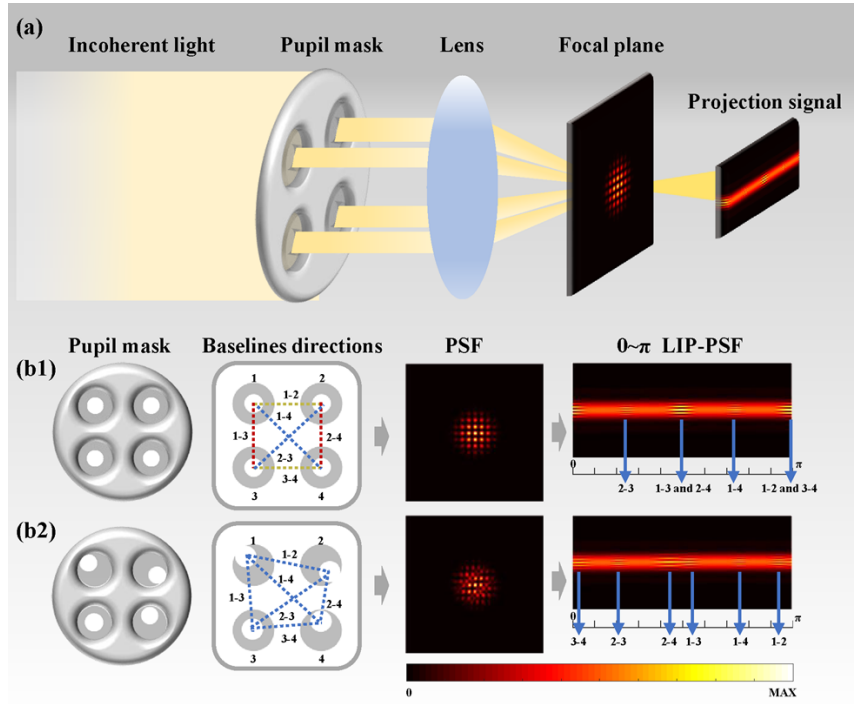

Fig. 3 Piston sensing implementation scheme. (a) Optical design of the piston sensing system. The mask without a parallel baseline is set on the exit-pupil plane to sample the wave reflected by the wavefront compensator. (b1) Interference fringes are coupled in parallel baselines. (b2) Interference fringes decoupled at different baseline orientations. With the same four subaperture imaging systems, a special mask was designed to verify the decoupling effect on the interference fringe. Under these conditions, the 1-2, 1-3, and 1-4 subaperture pairs have different interference paths in the LPI image.

We then devise a reference subaperture piston scanning technique that enables both a large range and high precision in piston sensing. The entire phasing process is shown in Fig. 4. First, the mask without parallel baselines is designed according to the position of the selected reference aperture, and the integral paths $(\theta_1, \theta_2......\theta_{n-1})$, which are determined by the baseline direction between the reference and remaining subapertures, can be confirmed. Subsequently, we control the piston actuator of the reference segment to traverse within the preset scanning range. At each scanning step, by applying n-1 times the LPI and irrelevant variable elimination to the PSF image, the n-1 sets of interference fringes can be extracted from the integral signal, and the intensity of the central point of the decoupled fringe can be simultaneously calculated and recorded as metric function vectors. In principle, the cophasing capture range provided by piston-sweep phasing is limited only by the dynamic range of the piston actuator. With the traverse ending, the scanning results can be expressed as:

$$\mathbf{Metrics} = [\ \mathbf{IMP}_1^T(x),\ \mathbf{IMP}_2^T(x),\ ...\ ,\ \mathbf{IMP}_i^T(x)\ ] \quad i = 1,2....n \quad (17)$$

where x represents the scanning position of the reference piston actuator, $\mathbf{IMP}_i$ is the vector of intensity of the main peak corresponding to each decoupled fringe, and n is the number of nonreference subapertures. The highest contrast and main peak intensity are achieved when the pistons between the subaperture and the reference aperture to the focal plane are zero. Consequently, the x value that corresponds to the maximum $\mathbf{IMP}_i$ is indicative of pistons between the reference and $i$th nonreference subapertures. The piston sensing results can be expressed as follows:

$$\mathbf{Pistons} = [\ -x_1(\mathbf{IMP}_{1\max}{}^T),\ -x_2(\mathbf{IMP}_{2\max}{}^T),\ \ldots,\ -x_i(\mathbf{IMP}_{i\max}{}^T)\ ]\quad i=1,2\ldots n \quad (18)$$

Compared with the other PSF-based techniques, the proposed method has the following advantages. First, it has a large-scale detection range, which is limited only by the dynamic range of the piston actuator. Second, in contrast to image quality optimization schemes, such as simulated annealing [52] and stochastic parallel gradient descent [53] algorithms, our approach decouples the fringes of each subaperture pair directly from the PSF image and creates a metric vector to simultaneously perform piston sensing; consequently, the decrease in detection accuracy caused by metric coupling can be eliminated. Finally, distinct from maximizing the image spatial frequency using the modulation transfer function (MTF, the amplitude of the PSF's Fourier transform) metric [54], our method decouples the fringes between subaperture pairs directly from the PSF image, thereby avoiding occasional instances of whole-wave errors due to the influence of noise on the extremal points in the MTF metric.

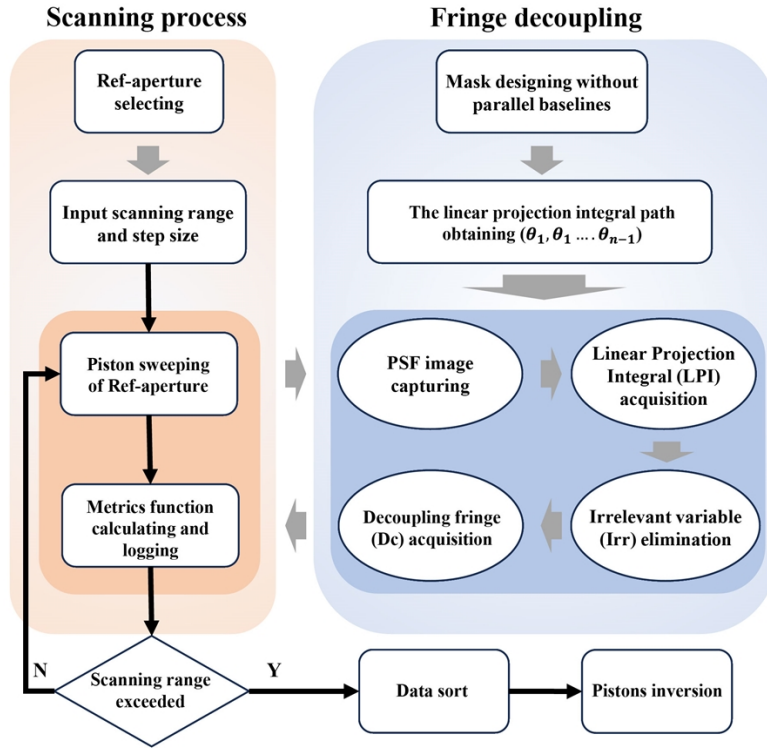

Fig. 4 Flowchart of the reference subaperture piston scanning cophasing technique based on LPI-based fringe disentanglement.

To validate the feasibility of the proposed method discussed in Section 2, we conducted one simulation and two experiments on different optical platforms.

### 3.2 Simulation Results

In numerical simulations, a synthetic aperture array with eight subapertures was used to verify the potential of the LPI, especially in terms of decoupling properties and the accuracy of piston sensing. The imaging system consists of eight subapertures with diameters of 20 mm, which are uniformly distributed within the circular ring with a 90 mm diameter in the outer circle. The broadband light with a wavelength of 550-650 nm passes through the pupil plane. The equivalent focal width is 2000 mm, and the pixel size of the detector is 3.45 μm. To simulate the real scene, 1% Gaussian noise (variance $\sigma^2 = 1\%$ of that of signal values) is added to each intensity image recorded by the CCD. As mentioned in Section 3.1, a mask without a parallel baseline is placed on the exit-pupil plane to eliminate the coupling relationship of interference fringes, as shown in Fig. 5(a). The mask is composed of eight subapertures with a diameter of 10 mm, and the angles between the direction vectors of each subaperture and the positive x-axis are 7.5°, 52.5°, 90°, 127.5°, 172.5°, 217.5°, 270° and 322.5°. Subaperture 3 is used as a reference. According to the mask distribution, seven projection integration paths corresponding to seven groups of decoupled interference fringes can be determined: 48.75°, 63.75°, 108.75°, 131.25°, 153.75°, 0° and 46.25°.

Figs. 5(b1) - 6(b3) show the PSF and its LPI image (0°-180°) in the system calibration, piston loading and piston correction states, respectively. Seven projection integral paths are marked on each of the LPI images, as shown in Figs. 5(b4) - (b6), respectively. In the piston loading process, subaperture 3 is taken as the reference, and the relative pistons of the remaining subapertures are 800, 1100, -2100, 400, -900, 3200, and -1600 nm. Because each subaperture beam interferes with each other on the detector plane, the fringe information between the remaining subapertures and the reference aperture is completely submerged, as shown in Fig. 5(b2). Moreover, the fringe contrast of seven integral curves in the corresponding LPI image, which is shown in Fig. 6(b5), becomes blurry, and the main peak of the decoupled fringes is much smaller than the calibration values in Fig. 6(b4). In piston sensing process, we command the reference aperture's piston position to scan from -10 μm to 10 μm with a step of 10 nm, and the normalized central intensities of seven integral curves which displays a unique maximum intensity in the whole scanning range are illustrated in Fig. 6(c). Consequently, the shift optical path length of reference aperture corresponding to the seven maximum intensity is the relative pistons of each subaperture. With piston compensation, the eight subaperture beams obviously interfere with each other, generating the high-brightness main peak shown in Fig. 6(b3). The LPI image of the corrected PSF is shown in Fig. 6(b6), in which the central intensities of seven integral curves are recovered. The correction residuals are shown in Table 1, and the average wavefront RMS is 0.022 wave at the 600 nm central wavelength. In order to verify the cophasing capture range is limited only by the dynamic range of the piston actuator, millimeter-level capture range co-phasing process has been simulated and the simulation results are shown in Fig. S1 and Table S1 of Supplement 1.

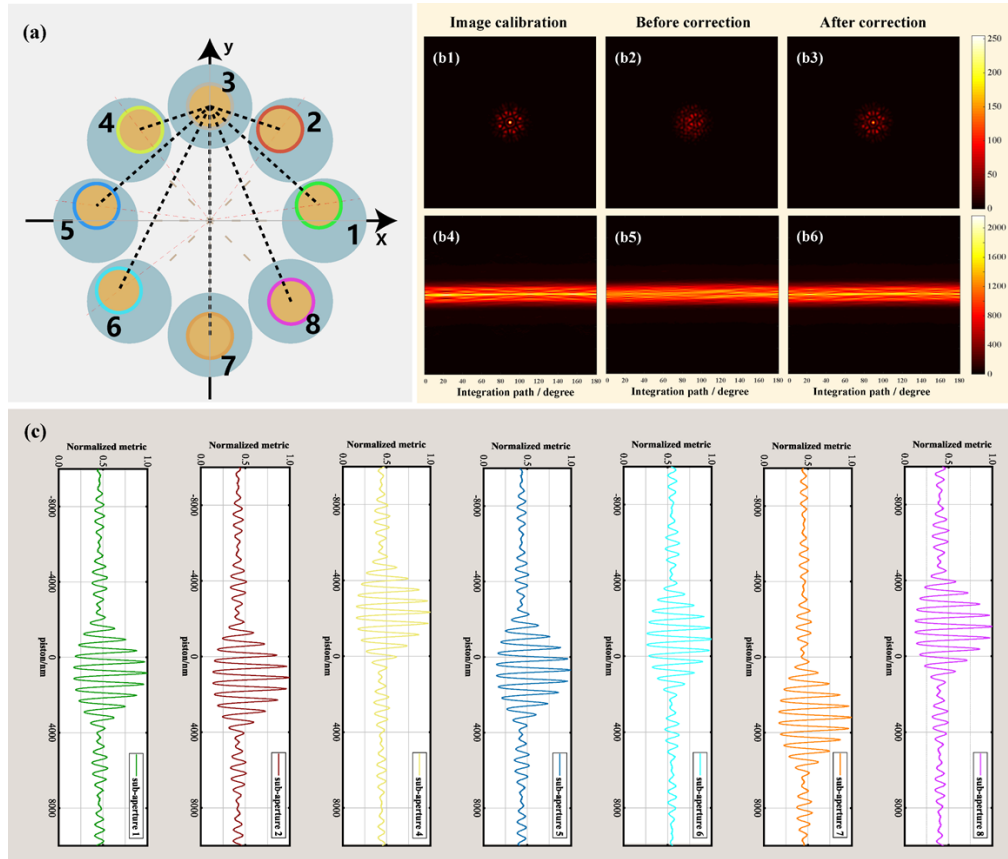

Fig. 5 Piston sensing simulation results. (a) The exit–pupil distribution of the simulated synthetic aperture imaging system consists of eight subapertures with diameters of 20 mm, which are uniformly distributed within the circular ring with a 90 mm diameter in the outer circle. (b) PSF images and its LPI images in the states of system calibration, piston loading and piston correction. (c) While scanning the reference aperture's piston position from -10 μm to 10 μm with a step of 10 nm, the normalized central intensities of seven integral curves are simultaneously recorded. The positions corresponding to each maximum magnitude within the entire scanning curve represent the pistons between the reference and nonreference subapertures.

**Table 1 Residual pistons of each nonreference aperture before and after correction**

| Subaperture | Sub1 | Sub2 | Sub4 | Sub5 | Sub6 | Sub7 | Sub8 | Wavefront/ RMS |
|---|---|---|---|---|---|---|---|---|
| Load piston/nm | 800 | 1100 | -2300 | 700 | -900 | 3200 | -1600 | 2.716λ |
| Residual/nm | -20 | -10 | 20 | 10 | 0 | 0 | 20 | 0.022λ |

### 3.3 Optical Desktop Experimental Results

An optical desktop experiment is implemented in a simplified synthetic aperture setup, as shown in Fig. 6(a). The piston actuators are constructed by a wavefront compensator composed of eight piezoelectric Fast Steering Mirrors (FSM), which have the same aperture distribution as the synthetic aperture array in the numerical simulation. Upon applying a voltage to the actuator, the FSM generates piston motion, thus introducing a piston error in the light beam of the corresponding subaperture. In this experiment, a 550 nm-650 nm broadband light source is transmitted to a collimator, and an expanded parallel light beam is subsequently formed. Reflected by a wavefront compensator placed at a 45° angle, the light passes through the multi-aperture mask, where there exists a one-to-one relationship between the submirrors and the corresponding subapertures. The pupil arrangement is also the same as that of the mask mentioned in Section 3.2, and limited by experimental devices, subapertures 1, 3, 4, 5, 6 and 8 were chosen for the experiment. Then, the light beams from the pupil are concentrated by using a lens with 2 m focal length. To evaluate the accuracy of the piston calibration, the converging beam is split into two separate light beams. In the transmission path, the light beam is directly captured by a phasing camera. In the reflection path, a diaphragm with multiple filters is placed in front of the calibration camera. When we rotate the diaphragm, the polychromatic light can be transformed to specific monochromatic light by different filters. The piston calibration is considered complete, provided that the PSF image in the calibration camera remains stable despite variations in the filter.

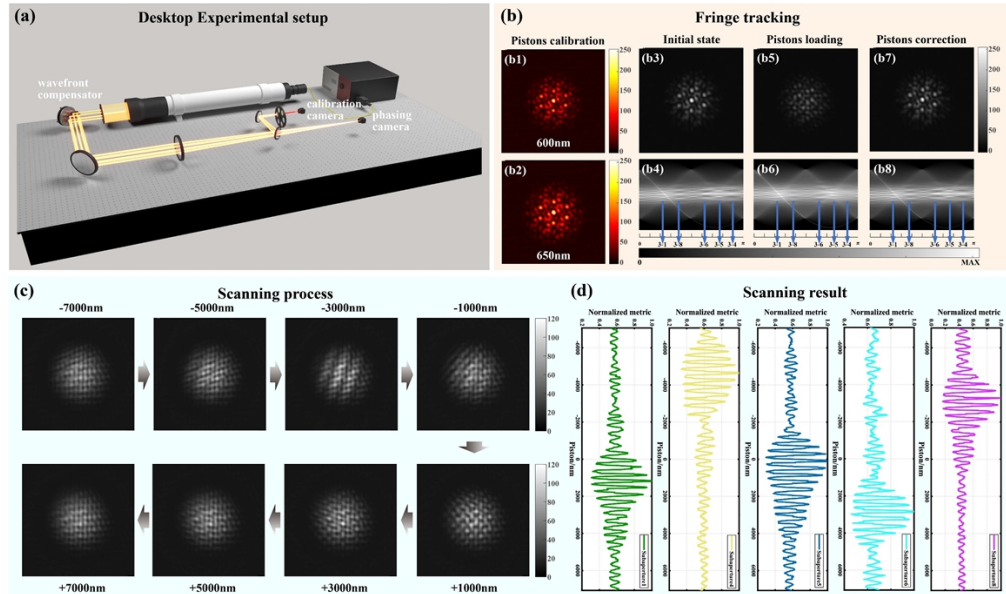

Fig. 6 Optical desktop experimental results. (a) Desktop experimental setup. A 550 nm-650 nm broadband light source is transmitted to a collimator and then reflected by a wavefront compensator, which is placed at a 45° angle. The light beams from the pupil are concentrated by using a lens with 2 m focal length and finally split into the phasing camera and the calibration camera. (b) Fringe tracking process. (b1) and (b2) show the piston calibration results of the imaging system. (b3) shows the PSF image after system calibration, and (b5) and (b7) show the PSF images before and after piston correction, respectively. (b4), (b6) and (b7) are the corresponding LPI images. (c) The reference subaperture piston scanning process. (d) Scanning results.

Optical desktop piston sensing experimental results are shown in Fig. 6(b), 6(c) and 6(d). Above all, a 600 nm filter and a 650 nm filter are selected to calibrate the residual pistons of the optical path until the optimal PSF image can be simultaneously captured in these two filter states, as illustrated in Fig. 6(b1) and 6(b2), respectively. With system calibration completed, the broadband PSF image captured in the phasing camera and its LPI image are shown in Fig. 6(b3) and 6(b4), respectively. Since the selected reference aperture and the design of pupil mask are exactly consistent with numerical simulation, the five integral paths corresponding to five groups of decouple interference fringes can be determined, which are marked on each LPI image. Subsequently, a set of pistons are sequentially loaded to the nonreference subapertures. The degenerated PSF image and its LPI image are shown in Fig. 6(b5) and 6(b6), respectively.

**Table 2 Residual pistons of each nonreference subaperture before and after correction**

| Subaperture | Sub1 | Sub4 | Sub5 | Sub6 | Sub8 | Wavefront/ RMS |
|---|---|---|---|---|---|---|
| Load piston/nm | 1131 | -4623 | 113 | 2813 | -3273 | 4.789λ |
| Residual/nm | 1159 | -4637 | 98 | 2828 | -3294 | 0.032λ |

In the piston sensing process, the piston actuator of the reference subaperture scans uniformly from -7 μm to 7 μm with a step of 10 nm, and the optimization evolutions of five criterion functions are presented in Fig. 6(d). The piston sweeping procedure is briefly depicted in Fig. 6(c). Considering the green curve in Fig. 6(d), the peak value of the curve is attained when the piston actuator of reference subaperture moves to 1131 nm position, which indicates the piston value between the reference subaperture and subaperture 1. Consequently, we can simultaneously obtain the five predicted pistons of which the inversion operations are implemented to realize the cophasing of the multiple subapertures. The correction residuals are shown in Table 2, and the RMS value of the wavefront is 0.032λ. The PSF after piston compensation and its LPI image are shown in Fig. 6(b7) and Fig. 6(b8), respectively. The experimental results show a conformation between the corrected interference pattern and the ideal distribution after calibration, which clearly demonstrates the effectiveness of the proposed method.

### 3.4 1.5-meter Flat Imaging System Experimental Results

The other experiment was implemented in the 1.5 m segmented flat imaging system [55], which is composed of four parts: a 1.5 m collimator, primary mirror, achromatic unit, and aft optics module. As illustrated in Fig. 7(a), with an NKT supercontinuum laser mounted at the focal point of a 1.5 m collimator, the parallel light passes through the eight flat elements and chromatic aberration correction module and then collimates into the aft optics module, which provides real-time alignment and phasing. The primary mirror, which has the same aperture distribution as the wavefront compensator, consists of eight diffractive elements with a diameter of 352 mm, and subapertures 2, 3, 4, 6 and 8 are chosen for the system experiment, while subaperture 3 is selected for reference.

In the engineering-level experiment, due to the influence of air turbulence, ground vibration and truss oscillation, the beams could be susceptibly disturbed during long-distance transmission, thus making multiple spots shake randomly in the phasing camera. Consequently, a high-accuracy line-of-sight jitter rejection is a prerequisite for the cophasing of a flat imaging system. We therefore use the two-order alignment scheme to stabilize the optical axes, in which the first order restrains the high-frequency holistic vibration and the second order compensates for the subapertures' respective low-frequency drifts. With the alignment completed, the five focused spots are stably overlapped in the center of the image plane of the phasing camera, as presented in Fig. 7(b1).

Then, the phasing process is started by applying the LPI-based piston sensing technique. Due to the machining tolerances and assembling inaccuracies of the primary mirror, the cumulative pistons between each flat element are generated, thus leading to distortion of the PSF image. The PSF image and its LPI image under the initial state are shown in Fig. 7(b1) and (b2), respectively. During the phasing process, consistent with the baseline direction between the reference and nonreference subapertures, four projection integral paths are first confirmed, and the four metric vectors, which respond only to the piston, are subsequently constructed. Second, we command the piston actuator of the reference subaperture to scan uniformly in the whole range from -5.6 μm to 5.6 μm with a step of 10 nm. At each scanning step, the confirmed integral paths are utilized to perform LPI operations on the PSF which are captured on the phasing camera in Fig. 7(a1), and the corresponding intensity integral indices are simultaneously recorded in the metric vectors. The scanning process is illustrated in Fig. 7(c). Finally, the negative value of piston actuator's position, which corresponds to the maximum in each metric vector, is regarded as the relative piston positions and is then applied to the other piston actuators of nonreference subapertures. The whole close-loop process is displayed in Visualization 1. The PSF image after compensation and its LPI image are presented in Fig. 7(b3) and (b4), respectively. The scanning result is shown in Fig. 7(c), which contains the variation in all metric indices during piston sweeping.

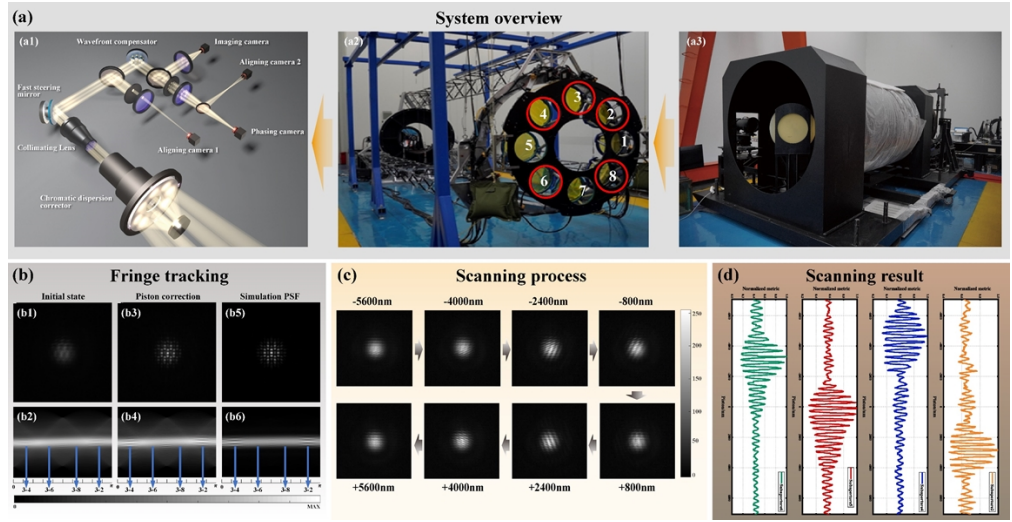

Fig. 7 1.5-meter flat imaging system experimental results. (a) System overview. With an NKT supercontinuum laser mounted at the focal point of a 1.5 m collimator, the parallel light passes through the eight flat elements and chromatic aberration correction module and then collimates into the aft optics module, which provides real-time alignment and phasing. (b) Fringe tracking process. (b1) - (b4) show the PSF image and its' LPI image under initial state and piston corrected. (b5) and (b6) show the calibration simulation image. (c) The reference subaperture piston scanning process. (d) Scanning results.

So far, we have proven the potential of the LPI-based piston sensing technique through numerical simulation, desktop experiments in the laboratory, and engineering experiment in ground testbed. Among them, in ground-based engineering experiments, the constantly variations in optical path length caused by external disturbance makes the main peak of the corrected PSF image disappeared. Consequently, with the piston compensation done, the analytical piston sensing method [55] is adopted, which guarantees phase locking of the system.

## 4. Application II - Parallel Visibility Measurement

Since the LPI-based fringe disentanglement technique is able to extract the interference signal of each baseline from PSF image, it is a naturally suitable choice for object visibility function measurements in stellar interferometry. The conventional ground-based interferometer adjusts the baseline direction through the rotation of the earth and subsequently utilizes the beam combination (BC) device to extract the interference information between each pair of subapertures, which is referred to as the pairwise combination scheme. In practice, such pairwise combination schemes are limited by the optomechanical stability of the beam combiners, which probably affects the stability of the relationships between the fringe positions of different baselines []. Based on the LPI operation, the visibility of each individual baseline can be directly distinguished from the mixed interference pattern, which eliminates the dependence on the optical splitter, referred to as all-in-one scheme. In this section, both simulations and experiments are conducted to verify the feasibility of the proposed method.

*4.1 Simulation results*

Simulations are performed with an extended target, as shown in Fig. 8(a). In the stellar interferometry imaging system, when the field of view is reduced to the half-width of the fringe package, much smaller than $PSF_0$ such that the speckle pattern can be displayed as a convolution of the object brightness distribution $O(x,y)$ with the fringe pattern. After calibrating the wavefront through the observation of separated referenced stars, the image intensity distribution can be expressed as follows:

$$\begin{aligned} I(x,y,\lambda) &= PSF_0(x,y,\lambda)\left\{O(x,y)\otimes\left\{n+2\sum_{j=1}^{n-1}\sum_{k=j+1}^{n}\cos\left[\left(\mu_{p_j}-\mu_{p_k}\right)\frac{2\pi x}{\lambda f}+\left(v_{p_j}-v_{p_k}\right)\frac{2\pi y}{\lambda f}\right]\right\}\right\} \\ &= CI_0\left\{n+2\sum_{j=1}^{n-1}\sum_{k=j+1}^{n}\left|\mu(R_{j,k})\right|\cos\left[\phi(R_{j,k})-\left(\mu_{p_j}-\mu_{p_k}\right)\frac{2\pi x}{\lambda f}-\left(v_{p_j}-v_{p_k}\right)\frac{2\pi y}{\lambda f}\right]\right\} \end{aligned} \quad (19)$$

in the first line in the notation as convolution, and in the second line as the sum of products of the visibility function with the cosine functions at individual spatial frequencies $R_{j,k}$ which is determined by the entrance pupil distribution. $\mu(R_{j,k})$ is the amplitude of the visibility function, and $\phi(R_{j,k})$ is the phase of the visibility function. In the simulation, the spatial frequency spectrum of the object can be obtained by applying the Fourier transform, as shown in Fig. 8(b). Accordingly, the amplitude and phase of the object visibility function can be respectively obtained by feeding the real part and imaginary part of the frequency spectrum into the modular operation and arctangent function. The exit pupil distribution is illustrated in Fig. 8(c), which determines the PSF of the imaging system.

Based on both the conventional pairwise combination scheme and the proposed all-in-one scheme, the object visibility extraction results corresponding to three baselines with different lengths and orientations are presented in Figs. 8(d) and 8(e), respectively. In the pairwise combination scheme, the selected baselines are shown in Figs. 8(d1), 8(d4) and 8(d7). The corresponding interference fringes are exhibited in Figs. 8(d2), 8(d5) and 8(d8). According to the intensity distribution of the fringes, the visibility function can be derived by the ABCD method, which relies on measuring the intensity at four different points on an individual fringe that are separated by 1/4 of the fringe spacing, as shown in Figs. 8(d3) - 8(d6) and 8(d9).

In the all-in-one scheme, the speckle pattern simultaneously combining the interference fringes between all sub-apertures is shown in Fig, 8(e2). The LPI-based fringe disentanglement technique is subsequently used to decouple the fringe of each baseline directly from the speckle pattern. The integral image is shown in Fig. 8(e3). Based on the exit–pupil distribution, three projection integral paths corresponding to the baseline orientations can be confirmed, along which three integral signals containing the interference information from each baseline can be extracted, as shown in Figs. 8(e4), 8(e6), and 8(e8), respectively. After the elimination of irrelevant terms, three interference fringes are analytically disentangled, as shown in Figs. 8(e5), 8(e7), and 8(e9). The spatial complex visibility function measurement results are shown in Table 3, which the amplitudes of each digitally decoupled visibility function and the corresponding phases are much close to the real values and pairwise measurement results. The fringe digitally disentanglement and spatial complex visibility function measurement results of remaining baselines are respectively shown in Fig. S2 and Table S2 of Supplement 1.

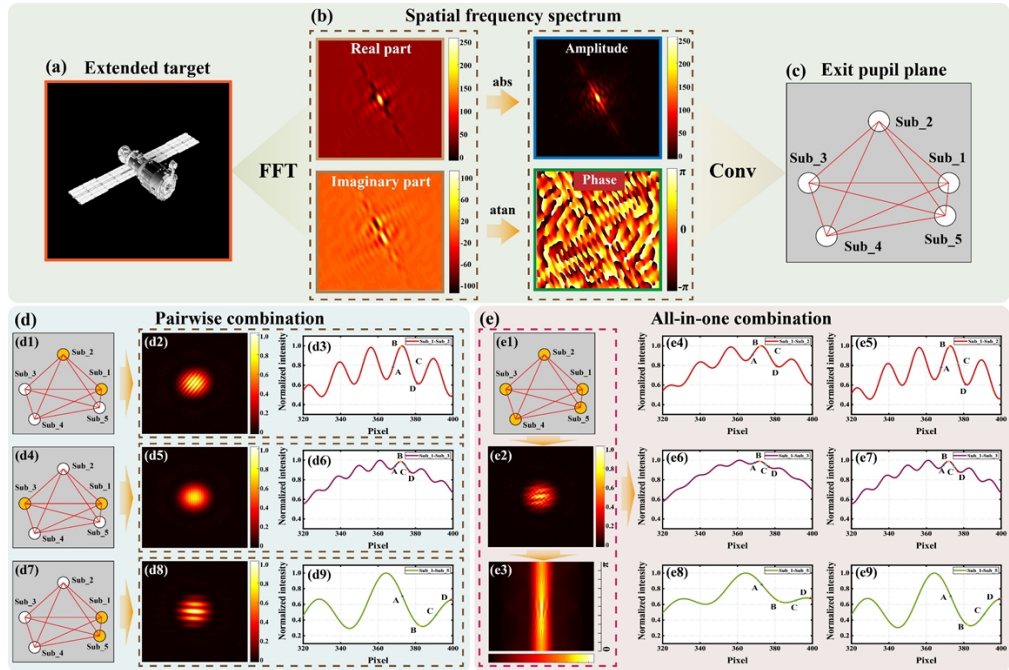

Fig. 8 Extend target fringe digitally disentanglement simulation results.

Table 3 Extend target spatial complex visibility function parallel measurement simulation results

| Beam combination scheme | Baseline | | | | | |
|---|---|---|---|---|---|---|
| | Sub_1-Sub_2 | Sub_1-Sub_3 | Sub_1-Sub_5 | Sub_1-Sub_2 | Sub_1-Sub_3 | Sub_1-Sub_5 |
| | Amplitude | | | Phase | | |
| Object visibility function | 0.2218 | 0.0496 | 0.5362 | 2.8819 | 2.9293 | 0.0974 |
| Pairwise combination scheme | 0.2317 | 0.0509 | 0.5224 | 3.0563 | 3.0128 | 0.1043 |
| All-in-one scheme (decoupled) | 0.2344 | 0.0516 | 0.5153 | 3.0635 | 2.9204 | 0.1052 |

## 4.2 Experimental results

Utilizing the fringe digitally disentanglement theory, the optical desktop experiments for point target spatial complex visibility function parallel measurement has been set up. The exit-pupil distribution is consistent with the simulation parameters in Section 4.1. The interference fringes corresponding to each baseline were recorded in both pair-wise combination and all-in-one measurement scheme, and three groups of these fringes were compared, as illustrated in Fig. 9. The equivalent focal length of imaging system is 2 m, the pixel size of the detector is 1.67 μm, and the wavelength of monochromatic illumination is set to 639 nm.

In the pairwise combination scheme, both the baseline distribution and the interference fringes of three pair of subapertures are respectively shown in Figs. 9(a1) - 9(a9). In all-in-one measurement scheme, the speckle pattern via image-plane beam combinations and its' LPI image are shown in Figs. 9(b2) and 9(b3). Three integral signals containing the interference information from speckle pattern can be extracted, and subsequently the corresponding interference fringes are analytically disentangled, as shown in Figs. 9(b4) -9(b9). It is apparent that the intensity distribution of each digitally decoupled fringe exhibits a great similarity to that of the interference fringe obtained by pairwise combination.

The spatial complex visibility function measurement results are shown in Table 4. In the pairwise combination scheme, affected by noise from the CCD, the amplitudes of each visibility function are 0.9617, 0.9828, and 0.9943 and the corresponding phases are -0.0117, -0.0462 and 0.0034. In the all-in-one scheme, the visibility function of digitally disentangled fringes is much close to the calibration value and pairwise measurement results, of which the amplitude and phase have absolute measurement errors of 3.02% and 2.60% from their ideal values. The point target fringe digitally disentanglement and corresponding spatial complex visibility function measurement results of remaining baselines are respectively shown in Fig. S3 and Table S3 of Supplement 1.

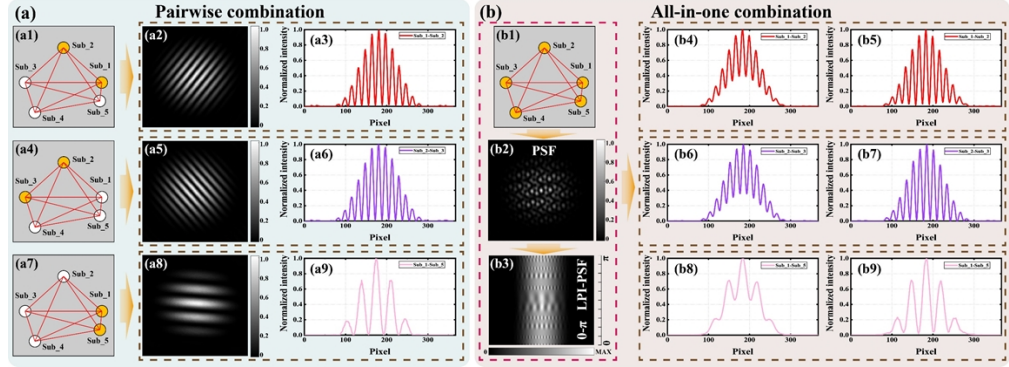

Fig. 9 Point target fringe digitally disentanglement experimental results.

**Table 4 Point target spatial complex visibility function parallel measurement experimental results**

| Beam combination scheme | Baseline | | | | | |
|---|---|---|---|---|---|---|
|  | Sub_1-Sub_2 | Sub_2-Sub_3 | Sub_1-Sub_5 | Sub_1-Sub_2 | Sub_2-Sub_3 | Sub_1-Sub_5 |
|  | Amplitude | | | Phase | | |
| Object visibility function | 1 | 1 | 1 | 0 | 0 | 0 |
| Pairwise combination scheme | 0.9617 | 0.9828 | 0.9943 | -0.0117 | -0.0462 | 0.0034 |
| All-in-one scheme (decoupled) | 0.9508 | 0.9777 | 0.9809 | -0.0133 | -0.0609 | 0.0039 |

## 5. Conclusion

For optical imaging interferometers consisting of more than two apertures, combining all the sub-aperture beams to the same focus results in spatial accumulation of each set of fringes of each pupil pair on the image plane. The speckle-like interference pattern losing fringe contrast, spacing and direction brings problems in information acquisition. To address this issue, we establish a theoretical model of fringe disentanglement based on linear projection integration through rigorous mathematical derivation, which is capable of digitally extracting pairwise fringes from a mixed interference pattern of many sub-apertures. The accuracy of the formulas we derive is verified by numerical simulation. Two applications using this fringe disentanglement theory have been presented. One is piston sensing. The theory-based phasing technique is applicable to multiple apertures without any additional optical elements and has high accuracy with long capture range that only depends on the active actuator range. We utilize the method to phase our 1.5-meter flat telescope, demonstrating the validity for engineering implementation. The other is parallel visibility measurement. The proposed fringe disentanglement theory offers an alternative solution to determine the individual visibility in spatial domain for all-in-one combination, which can effectively reduce crosstalk and has higher computational efficiency.

This theory provides new theoretical insights for acquisition of interferometric information and enables advanced applications for imaging interferometers. Meanwhile, this theory may not only benefit optical imaging but also interferometry-based measurements, by providing an possibility of realizing multiple channel measurement based on an interferometer with more than two interference arms.

**Funding.** National Key Research and Development Program of China (2022YFB3901900); National Natural Science Foundation of China (62175243); Youth Innovation Promotion Association, CAS (2020372); Outstanding Scientist Project of Tianfu Qingcheng Program

**Disclosures.** The authors declare no conflicts of interest.

**Data availability.** Data underlying the results presented in this paper are not publicly available at this time but may be obtained from the authors upon reasonable request.

**Supplemental document.** See Supplement 1 and Visualization 1 for supporting content.

## References

1. B. Kloppenborg, R. Stencel, J. Monnier, "Infrared images of the transiting disk in the ε Aurigae system," Nature 464, 870–872 (2010).
2. R. M. Roettenbacher, J. D. Monnier, H. Korhonen, "No Sun-like dynamo on the active star ζ Andromedae from starspot asymmetry," Nature 533, 217-220 (2016).
3. K. Ohnaka, G. Weigelt, & Hofmann, KH, "Vigorous atmospheric motion in the red supergiant star Antares," Nature 548, 310–312 (2017).
4. Stefan Kraus, "A triple-star system with a misaligned and warped circumstellar disk shaped by disk tearing," Science 369,1233-1238(2020).
5. Rubab Amin, Rishi Maiti, Yaliang Gui, Can Suer, Mario Miscuglio, Elham Heidari, Ray T. Chen, Hamed Dalir, and Volker J. Sorger, "Sub-wavelength GHz-fast broadband ITO Mach–Zehnder modulator on silicon photonics," Optica 7, 333-335 (2020)
6. Armin Hochrainer, Mayukh Lahiri, Radek Lapkiewicz, Gabriela B. Lemos, and Anton Zeilinger, "Interference fringes controlled by noninterfering photons," Optica 4, 341-344 (2017)
7. Mbaye Diouf, Zixi Lin, Mitchell Harling, Krishangi Krishna, and Kimani C. Toussaint, "Interferometric phase stability from Gaussian and space–time light sheets," Optica 10, 1161-1164 (2023)
8. Glindemann, By Andreas, "Principles of Stellar Interferometry," (Astronomy And Astrophysics Library).
9. E. K. Hege, J. M. Beckers, P. A. Strittmatter, "Multiple mirror telescope as a phased array telescope," Appl. Opt. 24(16), 2565-2576(1985).
10. S. S. Vogt, S. L. Allen, B. C. Bigelow, "HIRES: the high-resolution echelle spectrometer on the Keck 10-m Telescope," International Society for Optics and Photonics, 1994.
11. I. Smail, D. W. Hogg, L. Yan, "Deep optical galaxy counts with the Keck telescope,". Astron. J Letters, 449(2), L105(1995).
12. P. A. Sabelhaus, J. Decker, "James Webb Space Telescope: Project Overview,". IEEE Aerospace and Electronic Systems Magazine, 22(7), 3-13(2007).
13. M. W. Mcelwain, L. D. Feinberg, R. A. Kimble, "Status of the James Webb Space Telescope mission,". Space Telescopes and Instrumentation 2020: Optical, Infrared, and Millimeter Wave. SPIE, 11443T (2020).
14. Laginja, Iva, J. F. Sauvage, L. M. Mugnier, "Wavefront tolerances of space-based segmented telescopes at very high contrast: Experimental validation," Astron. Astrophys 658: A84(2022).
15. C. Anthony. Cheetham, G. Peter. Tuthill, Anand Sivaramakrishnan, and James P. Lloyd, "Fizeau interferometric cophasing of segmented mirrors," Opt. Express 20, 29457-29471 (2012)
16. Yujia Zhao, Ai Zhou, Huiyong Guo, "An Integrated Fiber Michelson Interferometer Based on Twin-Core and Side-Hole Fibers for Multiparameter Sensing," J. Lightwave Technol. 36, 993-997 (2018)
17. Khaled Alzahrani, David Burton, Francis Lilley, "Absolute distance measurement with micrometer accuracy using a Michelson interferometer and the iterative synthetic wavelength principle," Opt. Express 20, 5658-5682 (2012)
18. Pengqian Yang, Stefan Hippler, Casey P. Deen, "Characterization of the transmitted near-infrared wavefront error for the GRAVITY/VLTI Coudé Infrared Adaptive Optics System," Opt. Express 21, 9069-9080 (2013)
19. R. I Anderson, A Mérand, P Kervella, "Investigating Cepheid Carinae's cycle-to-cycle variations via contemporaneous velocimetry and interferometry," Mon. Not. R. Astron. Soc. 455(2016).

20. P. Haguenauer, R. Abuter, L. Andolfato, "The Very Large Telescope Interferometer v2012+," Proceedings of SPIE - The International Society for Optical Engineering (2012).
21. E. K. Baines, M. P. Doellinger, E. W. Guenther, "SPECTROSCOPIC AND INTERFEROMETRIC MEASUREMENTS OF NINE K GIANT STARS," Astron. J. 3:152(2016).
22. A. Benoît, F. A. Pike, T. K. Sharma, et al., "Ultrafast laser inscription of asymmetric integrated waveguide 3 dB couplers for astronomical K-band interferometry at the CHARA Array," J. Opt. Soc. Am. B 38, 2455–2464 (2021)
23. Denis Mourard, Philippe Bério, Karine Perraut, et al. "SPICA, Stellar Parameters and Images with a Cophased Array: a 6T visible combiner for the CHARA array," J. Opt. Soc. Am. A 34, A37-A46 (2017)
24. G. R. Petrov, F. Malbet, G. Weigelt, et al. "AMBER, the near-infrared spectro-interferometric three-telescope VLTI instrument," Astron Astrophys. 464, 1–12 (2007)
25. A. Ghasempour et al., "Building the next-generation science camera for the Navy Optical Interferometer," Proc. SPIE 8445, 84450M (2012).
26. DJ. Mortimer, DF Buscher, MJ Creech-Eakman, et al. "First laboratory results from FOURIER, the initial science combiner at the MROI," Proc. SPIE 11446 (114460V) (2020).
27. Pierre Haguenauer, Jean-Philippe Berger, Karine Rousselet-Perraut, et al. "Integrated optics for astronomical interferometry. III. Optical validation of a planar optics two-telescope beam combiner," Appl. Opt. 39, 2130-2139 (2000)
28. M. Benisty, J.-P. Berger, L. Jocoul, et al. "An integrated optics beam combiner for the second generation VLTI instruments," Astron. Astrophys. 498, 601–613 (2009).
29. Abani Shankar Nayak, Thomas Poletti, Tarun Kumar Sharma, et al. "Chromatic response of a four-telescope integrated-optics discrete beam combiner at the astronomical L band" Opt. Express 28, 34346-34361 (2020)
30. K. Perraut, L. Jocou, and J. P. Berger et al., "Single-mode waveguides for GRAVITY - I. The cryogenic 4-telescope integrated optics beam combiner," Astron. Astrophys. 614, A70 (2018).
31. S. Lacour, M. Nowak, and J. Wang et al., "First direct detection of an exoplanet by optical interferometry - astrometry and K-band spectroscopy of HR 8799 e," Astron. Astrophys. 623, L11 (2019).
32. J. Tepper, L. Labadie, and R. Diener et al., "Integrated optics prototype beam combiner for long baseline interferometry in the L and M bands," Astron. Astrophys. 602, A66 (2017).
33. A. Arriola, S. Mukherjee, and D. Choudhury et al. "Ultrafast laser inscription of mid-IR directional couplers for stellar interferometry," Opt. Lett. 39, 4820–4822 (2014).
34. S. R. McArthur, J. Siliprandi, and D. G. MacLachlan et al., "Ultrafast laser inscription of efficient volume Bragg gratings deep in fused silica using active wavefront shaping," Opt. Mater. Express 12, 3589–3599 (2022).
35. M. Benisty, J.-P. Berger, and L. Jocou et al., "An integrated optics beam combiner for the second-generation VLTI-instruments," Astron. Astrophys. 498, 601–613 (2009).
36. Gravity Collaboration, R. Abuter, M. Accardo, et al., "First light for GRAVITY: Phase referencing optical interferometry for the Very Large Telescope Interferometer," Astron. Astrophys. 602, A94 (2017).
37. N. Anugu et al., "MIRC-X: a highly sensitive six-telescope interferometric imager at the CHARA Array," Astron. J. 160, 158 (2020)
38. C. Lanthermann et al., "Modeling the e-APD SAPHIRA/C-RED ONE camera at low flux level. An attempt to count photons in the near-infrared with the MIRC-X interferometric combiner," Astron. Astrophys. 625, A38 (2019).
39. T. Gardner et al., "ARMADA. I. Triple companions detected in B-type binaries α Del and ν Gem," Astron. J. 161, 40 (2021).
40. G. Chanan, M. Troy, F. Dekens, S. Michaels, J. Nelson, T. Mast, and D. Kirkman, "Phasing the mirror segments of the Keck telescopes: the broadband phasing algorithm," Appl. Opt. **37**(1), 140–155 (1998).
41. G. Chanan, C. Ohara, and M. Troy, "Phasing the mirror segments of the Keck Telescopes II: the narrowband phasing algorithm," Appl. Opt. 39, 4706–4714 (2000).
42. M. Albanese, A. Wirth, A. Jankevics, et al. "Verification of the James Webb Space Telescope coarse phase sensor using the Keck Telescope," Proc. SPIE 6265, 62650Z (2006).
43. F. Shi, D. C. Redding, J. J. Green, et al. "Performance of segmented mirror coarse phasing with a dispersed fringe sensor: modeling and simulations,". Proc. SPIE, 5487:897-908(2004).
44. A. D. Marcos, A. M. Brian, and Antonin H. Bouchez, "Dispersed fringe sensor for the Giant Magellan Telescope," Appl. Opt. 55, 539-547 (2016)
45. A. H. Bouchez, B. A. McLeod, D. S. Acton, et al. "The Giant Magellan Telescope phasing system," Proc. SPIE 8447, 84473S (2012).
46. Y. Li and S. Wang, "Optical phasing method based on scanning white-light interferometry for multi-aperture optical telescopes," Opt. Lett. 46, 793-796 (2021)
47. D. J. Mortimer, and D. F. Buscher, "Crosstalk in image plane beam combination for optical interferometers," Mon. Not. R. Astron. Soc. 511, 4619–4632 (2022).
48. Seyed Mohammad Hashemi Rafsanjani, Mohammad Mirhosseini, Omar S. Magaña-Loaiza, Bryan T. Gard, Richard Birrittella, B. E. Koltenbah, C. G. Parazzoli, Barbara A. Capron, Christopher C. Gerry, Jonathan P. Dowling, and Robert W. Boyd, "Quantum-enhanced interferometry with weak thermal light," Optica 4, 487-491 (2017)

49. Antti Hannonen, Henri Partanen, Aleksi Leinonen, Janne Heikkinen, Tommi K. Hakala, Ari T. Friberg, and Tero Setälä, "Measurement of the Pancharatnam–Berry phase in two-beam interference," Optica 7, 1435-1439 (2020)
50. Rubab Amin, Rishi Maiti, Yaliang Gui, Can Suer, Mario Miscuglio, Elham Heidari, Ray T. Chen, Hamed Dalir, and Volker J. Sorger, "Sub-wavelength GHz-fast broadband ITO Mach–Zehnder modulator on silicon photonics," Optica 7, 333-335 (2020)
51. Uihan Kim, Hailian Quan, Seung Hyeok Seok, Yongjin Sung, and Chulmin Joo, "Quantitative refractive index tomography of millimeter-scale objects using single-pixel wavefront sampling," Optica 9, 1073-1083 (2022)
52. I. Paykin, L. Yacobi, J. Adler, and E. N. Ribak, "Phasing a segmented telescope," Phys. Rev. E 91, 023302 (2015).
53. Z. Xie, H. Ma, X. He, B. Qi, G. Ren, L. Dong, and Y. Tan, "Adaptive piston correction of sparse aperture systems with stochastic parallel gradient descent algorithm," Opt. Express 26, 9541–9551 (2018).
54. W. Zhao, and Q. Zeng, "Simultaneous multi-piston measurement method in segmented telescopes," Opt. Express 25(20), 24540–24552 (2017).
55. Z. Xie, K. Yang, Y. Liu, T. Xu, B. Chen, X. Ma, Y. Ruan, H. Ma, J. Du, J. Bian, D. Liu, L. Wang, T. Tang, J. Yuan, G. Ren, B. Qi, and H. Yang, "1.5-m flat imaging system aligned and phased in real time," Photon. Res. 11, 1339-1353 (2023)